\documentclass[12pt]{article}
\usepackage{pudlatex}
\usepackage{fullpage}
\usepackage{amsmath}
\usepackage[all,cmtip]{xy}

\title{On matrices potentially useful for tree codes\\
{\normalsize\it preliminary version}}
\author{Pavel Pudl\'ak
\thanks{Institute of Mathematics, Czech Academy of Sciences, Prague. The author is supported by the project EPAC, funded by the Grant Agency of the Czech Republic under the grant agreement no. 19-27871X, and the institute grant RVO: 67985840.}}

\begin{document}

\maketitle

\begin{abstract}
Motivated by a concept studied in \cite{BCN}, we consider a property of matrices over finite fields that generalizes triangular totally nonsingular matrices to block matrices. We show that (1) matrices with this property suffice to construct good tree codes and (2) a random block-triangular matrix over a field of quadratic size satisfies this property. We will also show that a generalization of this randomized construction yields codes over quadratic size fields for which the sum of the rate and minimum relative distance gets arbitrarily close to~1.
\end{abstract}

\section{Introduction}

The problem of an explicit construction of good tree codes is an interesting longstanding open problem. In~\cite{P} we showed that a good tree code can be constructed from a triangular totally nonsingular matrix over a filed of polynomial size.  A \emph{triangular totally nonsingular matrix}%
\footnote{Throughout this note triangular means lower triangular.}
is a matrix $M$ such that every square submatrix of $M$ whose main diagonal is in the lower triangle (we call such submatrices \emph{admissible}) is nonsingular. Note that only those minors can be nonzero. Whether or not such matrices exist over polynomial size fields is an open problem.
In contrast, examples of such matrices over exponentially large fields and fields of characteristic zero are well-known. The triangular Pascal matrix is triangular totally nonsingular over rationals, in fact it is triangular totally positive. Since the minors are at most exponentially large, one can take the Pascal matrix modulo an exponentially large prime a get a triangular totally nonsingular matrix over the corresponding prime field. These matrices have been used in the so far best construction of tree codes~\cite{CHS}. In~\cite{BCN} Ben Yaakov, Cohen and Narayanan report on computer experiments that strongly suggest that subexponential primes do not suffice, which is also conjectured in~\cite{KL}. Therefore Ben Yaacov et. al~\cite{BCN} considered a weaker property that Pascal matrices over small finite fields could satisfy and proved that this property is sufficient for constructing good tree codes. They conjectured that Pascal matrices have this property over polynomial size fields and supported the conjecture by computing the smallest prime for several small instances of the Pascal matrix.

Our aim in this note is, first, to show that a property slightly weaker than the one considered in~\cite{BCN} suffices for constructing good tree codes and, second, to prove that a random matrix having a certain pattern of zero and nonzero elements over a field of quadratic size satisfies this property. Our point is that one should not focus only on the Pascal matrix. The Pascal matrix may have the required property but it may be very difficult to prove it. Once we know that there are other matrices that satisfy our weaker property, we can look for different constructions. E.g., constructions of totally nonsingular matrices over linear size fields are well-known and we may try to use them to construct matrices with the property that we propose here.

The concept that we propose in this note is a natural generalization of triangular totally nonsingular matrices to block matrices. We only need matrices consisting of blocks that are $2\times 1$ matrices (or, if you prefer, column vectors of length 2). So $(2,1)$-block triangular matrices have dimensions $2n\times n$. The property that we need is that all admissible $(2,1)$-block triangular submatrices have full rank. In this note we only show how to construct from these matrices tree codes with rate $1/3$ and minimum relative distance $1/2$ with \emph{polynomial size} input and output alphabets, but it is well-known that from any such code one can construct a tree code with \emph{constant size} alphabets and positive rate and minimum distance. 

The property studied in~\cite{BCN} can also be stated in terms of $(2,1)$-block triangular matrices. Ben Yaacov et al. take the matrix of odd columns of the triangular Pascal matrix. This matrix can be viewed as a $(2,1)$-block triangular matrix. They conjecture that in this matrix all admissible $(2,1)$-block triangular submatrices have a slightly stronger property than full rank, possibly with some small exceptions.

In Section~\ref{s-gen} we generalize the construction and show that there are linear codes over quadratic fields such that the sum of the rate and relative minimum distance is arbitrarily close to 1 (the Singleton bound). Whether or not one can actually achieve the bound 1 with codes over subexponential size fields remains an open problem.

In~\cite{KL} Karingula and Lovett studied a somewhat related topic. They estimated the probability that a random \emph{integer} matrix with entries in $\{-m\dts m\}$ is singular. They conjectured that for $m$ bounded from below by a sufficiently large polynomial in the dimension of the matrix, a random triangular integer matrix is triangular totally nonsingular. In the last section we propose a conjecture about certain random matrices that implies the conjecture of Karingula and Lovett.\footnote{After we finished a draft of this note we learned that Karingula and Lovett tried a similar approach.}

\section{Definitions and basic facts}

\subsection{Patterns}

We will call a \emph{pattern} any matrix with entries $0$ and $*$. We say that a matrix $M$ over a field is \emph{consistent with a pattern} $P$, if $M$ and $P$ have the same dimensions and $M$ has nonzero elements only where $P$ has stars. We will only consider \emph{left-bottom patterns} (abbreviated by lb-patterns) which are patterns in which every entry before or below an occurrence of $*$ is also $*$.  E.g., lower triangular matrices are matrices that are consistent with the \emph{triangular lb-patterns,} which are the lb-patterns consisting of a triangle of stars delimited by the main diagonal with the rest filled with zeros. Triangular lb-patterns are, by definition, square matrices. We will denote by $T_n$ the $n\times n$ triangular lb-pattern. Two more important examples are \emph{all-star patterns} and \emph{all-zero patterns} which are  matrices where all entries are stars, respectively zeros

A \emph{subpattern} $P'$ of a pattern $P$ is a submatrix of $P$; if $P$ is an lb-pattern, then, clearly, $P'$ is also an lb-pattern. 
We call an lb-pattern $P$ \emph{admissible}, if it is a square matrix and has stars on the main diagonal. 
\bfa
Let $P$ be a triangular lb-pattern and $P'=P[i_1\dts i_k|j_1\dts j_k]$ be the subpattern given by rows $i_1\dts i_k$ and columns $j_1\dts j_k$. Then $P'$ is admissible iff $i_1\geq j_1\dts i_k\geq j_k$.
\efa

\bfa\label{f2}
Let $P$ be a square lb-pattern with $*$ in the upper left corner. Then either $P$ is admissible, or $P$ can be decomposed into blocks
\[
\def\arraystretch{1.5}
P=
\left[\ba{c}
\ba{l|r}
A~&~O
\ea\\
\hline
Z
\ea\right]
\]
where $A$ is an admissible pattern and $O$ is a zero pattern.
\efa
\bprf
Let $P$ be an $n\times n$ lb-pattern with  $P_{1,1}=*$. If $P$ is not admissible, take the largest $m$ such that $P[1\dts m|1\dts m]$ is admissible. Such an $m$ exists, because the $P_{1,1}=*$. For this $m$, $P_{m+1,m+1}=0$, hence $P[1\dts m+1|m+1\dts n]$ is a zero matrix (which is a little bit more than what is stated in Fact).
\eprf

\subsection{Matrices}

Given a matrix $M$ and a pattern $P$ defined on rows $i_1\dts i_k$ and columns $j_1\dts j_l$, we will denote by $M[P]:=M[i_1\dts i_k|j_1\dts j_l]$.

We say that an $n\times n$ matrix $M$ is \emph{triangular totally nonsingular}, if for every admissible subpattern of the $T_n$ the corresponding submatrix of $M$ is nonsingular. Since only those square matrices that are induced by admissible patterns can be nonsingular, one often says ``all submatrices of $M$ that \emph{can be} nonsingular \emph{are} nonsingular''.

We will now generalize the introduced concepts to block matrices. An $(a,b)$-block matrix is a matrix $M$ whose entries are $a\times b$ matrices. If $M$ is an $m\times n$ $(a,b)$-block matrix, we can forget the blocks and view $M$ as an $am\times bn$ ordinary matrix; we will denote this matrix by $M^\flat$. We will say that the \emph{block dimensions of} $M$ are $m\times n$.

We say that a block matrix is consistent with a pattern $P$, if all blocks corresponding to zeros of $P$ are zero matrices. E.g., a lower triangular block matrix of dimension $n$ is a matrix is matrix consistent with $T_n$.

We want to generalize triangular totally nonsingular matrices to block matrices. In this note we only need it for matrices with blocks $2\times 1$. The generalization is straightforward except that we have to decide what to use instead of nonsingularity. We will use full rank.

\bdf
We say that a triangular $(2,1)$-block matrix $M$ is a \emph{totally full rank} block matrix, if for every admissible $(2,1)$-block submatrix $N$, $N^\flat$ is a full rank matrix.
\edf

In \cite{BCN} the authors considered a stronger condition: from each pair of consecutive rows (odd and even) of $N^\flat$ one can choose one row so that the resulting matrix is nonsingular.

\subsection{Linear tree codes}

We consider \emph{truncated tree codes}, which means that the code-words have some fixed finite length called the length of the code. We will omit the specification ``truncated'' in the sequel. 

Linear tree codes have generating matrices that are of the form $M^\flat$ for some $(a,b)$-block triangular matrix $M$ with $a>b$, see~\cite{P}. Here we consider lower triangular matrices, so the code is generated by the columns of $M^\flat$. In Section~\ref{s3} we will use $(3,1)$-block triangular generating matrices which define codes of rate $1/3$. We will view the code-words also as $(3,1)$-block matrices, or more precisely, $(3,1)$-block column vectors. The \emph{block length} of these vectors is the number of blocks.

Let $C$ be a code of block-length $n$. The \emph{minimum relative distance} is defined as follows. Take a nonzero vector $\vec c\in C$ and let $k$ be the index of the first nonzero block. The relative weight of $C$ is
\[
\min_{0\leq l\leq n-k}\frac{\mbox{the number of nonzero blocks between $k$ and $k+l$}}{l+1}. 
\]
The minimum relative distance of $C$ is the minimum of the relative weights of nonzero code-words.


\section{Two observations}\label{s3}

\bpr\label{p1}
Given a triangular totally full rank $(2,1)$-block matrix $M$ of block-dimension $n$ over a field $F$, one can construct a linear tree code $C$ 
over $F$ such that
\ben
\item $C$ has block length $n$;
\item the blocks of code-words have length $3$, thus the rate of the code is $1/3$;
\item the minimum relative distance is $>1/2$.
\een
\epr

\bprf
The proof uses an idea of~\cite{BH} by which Bhandari and Harsha proved that triangular totally nonsingular matrices can be used to construct tree codes that match the Singleton bound.%
\footnote{Our original proof in~\cite{P} was different and gave a stronger result. The difference is that we proved a bound on the relative number of nonzero \emph{entries} in a nonzero code word, which in general may be much smaller than the relative number of nonzero \emph{blocks}.}

Let $M$ be given. 
Let $N$ be a $(3,1)$-block triangular matrix obtained from $M$ by interleaving $M$ with the identity matrix. This means that we extend every $(2,1)$-block of $M$ to a $(3,1)$-block by adding a $0$ or $1$ at the top as follows. We add $1$s to blocks on the main diagonal and $0$s elsewhere.  
Consider the tree code generated by the columns of $N$.

The added 0s and 1s ensure that the coding function is 1-1. We only need to show the lower bound on the minimum distance. 
Let $\vec c_{j_1}\dts \vec c_{j_k}$ be columns of $N$ generating a nonzero code-word $\vec c$ (a column vector) using some nonzero coefficients. The condition that we have to verify depends on principal submatrices of $N$ which have the same properties as $N$. Therefore  
we may assume w.l.o.g. that $j_1=1$, i.e., $\vec c_{j_1}$ is the first column of $N$ and consider how many nonzero elements has the entire vector~$\vec c$.

We will consider two cases:

\medskip
1. $k> n/2$. Then we are done---the 1s of the identity matrix produce the $k$ nonzero blocks in $\vec c$.
 
\medskip
2. $k\leq n/2$. Let $S$ be the indices of the $(3\times 1)$-blocks of $\vec c$ that are zero. We need to show $|S|<n/2$. So 
suppose this is not the case. Then $|S|\geq n/2\geq k$. Consider the block submatrix $N[S|i_1\dots i_k]$. Let $K$ be this matrix with rows corresponding to the identity matrix omitted. 
Let $K'$ be the submatrix of $K$ with only the first $2k$ rows. So $K'$ is a $(2\times 1)$-block square submatrix of $M$.
$K'$ itself may not be admissible, but we know that the elements in the first column are nonzero and thus we may use 
Fact~\ref{f2}. According to this fact some left upper part of $K'$ is an admissible submatrix $K''$ and there are zeros to the right of it.
The columns of $K''$ sum to a zero vector, because the columns of $N[S|i_1\dots i_k]$  sum to a zero vector. 
But this is in contradiction with the assumption that $M$ is a triangular totally full rank $(2,1)$-block matrix. Hence $|S|<n/2$, which is the number of zero blocks in $\vec c$.
\eprf


\bpr\label{p2}
For every $n\geq 1$ and every field $F$, if $|F|\geq 2 n^2$, then there exists  a $(2,1)$-block-triangular totally full rank  matrix $M$  over $F$ of block-dimension $n$. 
Moreover, for finite fields,  if $|F|\geq\omega(n^2)$, then the probability that a random matrix has this property tends to~$1$ with $n\to\infty$.
\epr
{\it Proof.}
It is well-known that the triangular Pascal matrix is triangular totally nonsingular in~$\R$. From this, we get the desired matrix $M$ for fields of characteristic~0 by omitting even columns. So we only need to consider finite fields.

Let $F$ be a finite field and $q=|F|$.
We need to estimate probabilities of random matrices consistent with  $(2,1)$-block-admissible patterns. These are patterns that arise from admissible patterns by duplicating the entries to get $(2,1)$-blocks. In the rest of the proof, we will omit the specification $(2,1)$ and simply say {\it``block-admissible patterns''.} The reader should keep in mind that these patterns and matrices are now $2k\times k$ for some $k$.
We will denote by $E$ the elementary admissible pattern 
{\tiny$\left[\ba{c} *\\ *\ea\right]$.}
For a block-admissible pattern $P$ larger than $E$, we will denote by $P^-$ the pattern $P$ with the first two rows and the first column omitted; thus $P^-$ has the block-dimension one less than $P$. The pattern $P^-$ is, clearly, also block-admissible. We will denote by $\phi$ the property that a matrix has full rank.

 For a block-admissible pattern $P$ with some zeros, the probability that a random matrix consistent with $P$ does not have full rank may not be exponentially small; in fact, it may be as high as $q^{-2}$. Therefore we will use  conditional probabilities. 

\bcl\label{c-1}
Let $P$ be a block-admissible pattern larger than $E$. The probability that a random $2k\times k$ matrix $M$ consistent with $P$ does not have full rank under the condition that $M[P^-]$ has full rank is $q^{-k-1}$.
\ecl
\bprf
If the last $k-1$ columns are independent, then the matrix does not have full rank iff the first column is dependent on the last ones. This probability is equal to the probability that the first column is in the linear space spanned by the last ones, which is $q^{k-1}/q^{2k}$.
\eprf

\bcl
Let $T^2_n$ be the $2n\times n$ block-triangular pattern. Let $M$ be a random matrix over a finite field $F$ consistent with $T^2_n$. Then
the probability that for some block admissible subpattern $P$ of $T^2_n$, the matrix $M[P]$ does not have full rank is bounded by
\bel{e-cp}
\leq \sum_E\prob[\neg\phi(M[E])] + \sum_P\prob[\neg\phi(M[P])\wedge \phi(M[P^-])],
\ee
where the first sum is over all elementary admissible subpatterns $E$ of $T^2_n$ and the second sum is over all block-admissible subpatterns $P$ larger than $E$.
\ecl
\bprf
Suppose that there exists an admissible subpattern $P$ of $T^2_n$ such that $M[P]$ does not have full rank. Consider such a $P$ with the least dimension. Then $P$ is either $E$, or $P$ is larger than $E$ and $M[P^-]$ does not have full rank. Hence the bound follows by the union bound. 
\eprf

To prove the proposition, we need to show that for a finite field $F$ with $|F|\geq 2 n^2$, the probability that for some block admissible pattern $P$, the  matrix $M[P]$ is not full rank is less than 1. We observe that 
$\prob[\neg\phi(M[P])\wedge \phi(M[P'])]\leq\prob[\neg\phi(M[P])| \phi(M[P'])]$.
By Claim~\ref{c-1}, the latter probability is $q^{-k-1}$.
We also have $\prob[\neg\phi(M[E])]=q^{-2}$.
To bound the sum in~(\ref{e-cp}), 
we will estimate the number of block-admissible patterns of dimensions $2k\times k$ by $\leq\binom{n}{k}^2$ and use the assumption $q\geq 2n^2$. Thus we get
\[
\leq 
\sum_{k=1}^n\binom{n}{k}^2\cdot q^{-k-1}\leq
\sum_{k=1}^n \left(\frac{n^2}{q}\right)^{-k}\leq \sum_{k=1}^n 2^{-k}< 1.
\]
Clearly, 
the estimate goes to 0 when $q\geq\omega(n^2)$.
\qed


\section{A generalization}\label{s-gen}

We will now generalize the randomized construction from the previous section and show that the sum of the rate and relative minimum distance may be arbitrarily close to 1 for fields of quadratic size. To this end we need to consider more general linear codes.

Let $F$ be a field and let $s>t\geq 1$ be fixed constants. Let $N$ be an $(s,t)$-block triangular matrix of block dimension $n$ (thus $N^\flat$, which is $N$ with blocks forgotten, has dimensions $sn\times tn$). If $N$ satisfies a certain condition, we can use $N$ to define a tree code $C:(F^t)^n\to (F^s)^n$ as follows. For an input $\vec x\in (F^t)^n$ we take $\vec x^\flat$ (which is $\vec x$ with blocks forgotten), multiply it by $N^\flat$,  and then we introduce blocks of length $s$ on the vector $N^\flat \vec x^\flat$.
Deleting and introducing blocks may look as an unnecessary complication, but the point is that the input alphabet are vectors of length $t$ and the output alphabet are vectors of length $s$.

For the mapping $C$ to be a tree code, it is necessary and sufficient that the blocks on the diagonal of $N$ are full rank matrices. Further, one can always put the matrix into the \emph{normal form} in which blocks on the main diagonal are matrices starting with the $s\times s$ identity matrix and the blocks outside are matrices starting with $s\times s$ zero matrix. Thus in order to define such a code we only need a suitable $(s-t,t)$-block triangular matrix. The minimum distance will depend on the properties of this matrix; the rate is always~$t/s$. We will call codes defined above \emph{$(s,t)$-codes.}

\bpr
For every numbers $t$ and $n$ and every field $F$, $|F|\geq 2n^2$, there exists a linear $(2t+1,t)$-code
with rate $\frac t{2t+1}$ and minimum relative distance $\geq 1/2$. Thus 
the sum of the rate and relative minimum distance is $\geq 1-\frac 1{4t+2}$.
\epr
The bound on the size of the field does not depend on $t$, but note that the sizes of the input and output alphabets increase with $t$.

\bprf
The proof follows the lines of of the proof of Proposition~\ref{p2} with appropriate changes. First we will show that a random $(t+1,t)$-block triangular matrix over a field $F$, $|F|\geq 2n^2$ is totally full rank. To this end 
we need to estimate:
\ben
\item the probability that a random $(t+1)\times t$ matrix is nonsingular, and
\item for random $(t+1)k\times tk$ matrices $M$, where $k>1$, the conditional probability
  \[
\prob[M\mbox{ nonsingular }|\ M[t+2\dts (t+1)k|t+1\dts tk]\mbox{ nonsingular }].
  \]
\een
One can show using elementary calculations that the first probability is $<q^{-1}$ and the second one is $<q^{-k}$, where $q=|F|$. Hence the same computation as in the proof of Proposition~\ref{p2} shows that $M$ is $(t+1,t)$-block triangular totally nonsingular with positive probability.

From $M$ we get a generating matrix in the normal form by extending the $(t+1,t)$-blocks with zero and identity $t\times t$ matrices. The resulting $(2t+1,t)$-block matrix generates a tree code with rate $t/(2t+1)$. It remains to show that the minimum relative distance of this code is $\geq 1/2$. The proof of this bound is a literal translation of the proof of Proposition~\ref{p1}. 
\eprf

In order to see that the bound on the sum of the rate and minimum relative distance approaches the best possible value, we will check \emph{the Singleton bound for linear tree codes.} It does not hold exactly, only \emph{asymptotically}.

\bpr
For every $s>t>0$, there exists $c_{s,t}>0$ such that for every $n$ and every linear $(s,t)$-code of length $n$ the sum of the rate $r$ and the minimum relative distance $\delta$ satisfies
\[
r+\delta\leq 1+1/c_{s,t}n.
\]
\epr
\bprf
Let $M$ be a $(s,t)$ block matrix generating a code of block length~$n$. Assume that $M$ is in the normal form defined above. Take the first $k$ block columns of $M$, where $k$ will be determined later, and $k+l$ block rows where $l$ the largest integer such that $(s-t)l<kt$. From this $k+l\times k$ matrix we only take the bottom $l$ block rows and denote this matrix by $N$. 
If we forget the block structure, the resulting matrix $N^\flat$ has dimensions $sl\times tk$. In $N$ every $(s,t)$-block is off-diagonal, hence starts with $t$ rows of zeros. Thus  $N^\flat$ has $tl$ rows of zeros and $(s-t)l$ possibly nonzero rows. For we assume that $(s-t)l<kt$, there exists a nonzero vector $\vec x$ such that  $N^\flat \vec x=\vec 0$. Let $\vec y^{\ \flat}:=M^\flat \vec x$ and let $\vec y$ be the $s$-block vector obtained from $\vec y^{\ \flat}$ by introducing the blocks. Since $\vec x$ is nonzero, at least one of the first $k$ blocks of $\vec y$ are nonzero. On the other hand, the last $l$ blocks are zero. Hence the relative weight $\delta_{\vec y}$ of $\vec y$ is at most $k/(k+l)$. The condition $(s-t)l<kt$ is equivalent to $(s-t)l\leq kt-1$. Thus we have
\[
\delta_{\vec y}\leq\frac k{k+\frac{kt-1}{s-t}}=
\frac 1{1+\frac{t-\frac 1k}{s-t}}=\frac{s-t}{s-\frac 1k}.
\]
This gives the upper bound
\[
r+\delta\leq \frac ts + \frac{s-t}{s-\frac 1k}=
1+\frac{1-\frac ts}{sk-1}.
\]
One can easily check that the largest $k$ that we can take, which is the largest $k$ such that $k+l\leq n$, is a positive fraction of $n$.
\eprf


\section{Integer triangular totally nonsingular matrices}

While probabilistic arguments seem to fail to prove the existence of triangular totally nonsingular matrices over polynomially large fields, it is possible that they may be used for triangular totally nonsingular matrices over integers with polynomially large entries. Karingula and Lovett~\cite{KL} conjectured that for a sufficiently large polynomial $p(n)$, a random triangular integer matrix with entries in $[-m,m]$, $m\geq p(n)$, is with high probability totally nonsingular. (High probability only means $\geq 1-1/m^\alpha$ for some constant $\alpha>0$.) In this section we will state a conjecture that implies the conjecture of Karingula and Lovett.

An $n\times n$ lb-pattern $P$ is called \emph{reducible} if for some $k$, $1\leq k\leq n$, the subpattern $P[1,2\dts k|k+1,k+2\dts n]$ is an all-zero pattern; otherwise it is called \emph{irreducible}. Equivalently, an lb-pattern $P$ is irreducible iff $P_{i,i+1}=*$ for all $i=1\dts n-1$.

\begin{conjecture}
  There exists an $\eps>0$ and $c$ such that for every $n$, every $m\geq n^c$, and every $n\times n$ irreducible lb-pattern $P$, random matrices $M$ with entries in $[-m,m]$ consistent with $P$ satisfy
  \[
\prob[M\mbox{ is singular }|\ M[1,2\dts n-1|2,3\dts n]\mbox{ is nonsingular }] < m^{-\eps n}.
  \]
\end{conjecture}

If the conjecture is not true, it will most likely fail on the minimal irreducible lb-patterns. These are the patterns $P$ such that $P_{ij}=0$ iff $i+1<j$. For a matrix $M$ consistent with such a pattern, the condition that $M[1,2\dts n-1|2,3\dts n]$ is nonsingular is equivalent to the condition that $M_{i,i+1}\neq 0$ for all $1\leq i\leq n-1$, because then the subpattern $P[1,2\dts n-1|2,3\dts n]$ is triangular.

For an $n\times n$ pattern $P$, we will denote by $P^*:=P[1,2\dts n-1|2,3\dts n]$. If $P$ is irreducible, then $P^*$ is admissible. If $P$ is admissible, then the main diagonal can be covered by disjoint irreducible principal subpatterns $P_1\dts P_m$ (we allow $m=1$). We will call $P_1\dts P_m$ the \emph{decomposition} of $P$. We will use the following fact.

\bfa\label{f3}
Let $P$ be an admissible pattern and $P_1\dts P_m$ its decomposition. Then
a matrix $M$ consistent with $P$ is nonsingular iff all submatrices $M[P_1]\dts M[P_m]$ are nonsingular.
\efa

\bpr
If the conjecture above holds true, then for some $c$, there exist $n\times n$ triangular totally nonsingular integer matrices with entries in $[-n^c,n^c]$ for every $n\geq 1$. Moreover, if $c$ is sufficiently large, the probability that a random triangular matrix is totally nonsingular tends to $1$ with $n\to\infty$.
\epr
\bprf
The proof is based on the same idea as the one of Proposition~\ref{p2}.

By Fact~\ref{f3}, a triangular matrix $M$ is totally nonsingular iff for every irreducible subpattern $P$ of $T_n$, $M[P]$ is nonsingular. Suppose $M$ not totally nonsingular and let $P$ be an irreducible subpattern of $T_n$ of the smallest dimension such that $M[P]$ is singular. Then we there are two cases.
\ben
\item $P$ is a pattern with a single entry, and $M[P]=0$, or
  \item $P$ is larger. In this case let $P^*_1\dts P^*_m$ be the decomposition of $P^*$ into irreducible patterns. Since $P^*_1\dts P^*_m$ are smaller than $P$, $M[P^*_i]$ are nonsingular. Hence $P^*$ is also nonsingular. 
    \een
    Thus we can bound the probability that $M$ is not totally nonsingular by
\[    
\sum_{i\geq j}\prob[M_{i,j}=0] +
\sum_P\prob[M[P]\mbox{ is singular }\wedge\ M[P^*]\mbox{ is nonsingular }],
\]
where the second sum is over all irreducible subpatterns of $T_n$ of dimensions $\geq 2$. The rest of the proof is a computation similar to the one in the proof of Proposition~\ref{p2}, so we leave it to the reader.
\eprf



\begin{thebibliography}{1}

\bibitem{BCN}
I. Ben Yaacov, G. Cohen, and A. K. Narayanan:
{\it Candidate Tree Codes via Pascal Determinant Cubes.} Electron. Colloquium Comput. Complex. 27: 141 (2020).

\bibitem{BH}
S. Bhandari and P. Harsha: 
{\it A note on the explicit constructions of tree codes over polylogarithmic-sized alphabet.} arXiv:2002.08231 (2020)

\bibitem{CHS}
G. Cohen, B. Haeupler, and L. J. Schulman:
{\it Explicit binary tree codes with polylogarithmic size alphabet.}
 STOC 2018: 535-544

\bibitem{KL}
S. R. Karingula and S. Lovett: {\it Codes over integers, and the singularity of random matrices with large entries.} arXiv:2010.12081 (2020).

\bibitem{P}
P. Pudl\'ak: {\it Linear tree codes and the problem of explicit constructions.} 
Linear Algebra and its Applications 490, 2016, 124-144. 

\end{thebibliography}
\bye